\documentclass[prd,showpacs,nofootinbib,preprint, 11 pt]{revtex4}
\usepackage{amssymb}
\usepackage{amsmath,graphicx,color,epsfig}
\usepackage{pstricks}
\usepackage{float}

\begin{document}

\begin{flushright}
DESY 10-045\\
FTUAM-2010-09\\
May 2010
\end{flushright}

\title{Prospects of measuring the CKM matrix element \boldmath{$\vert V_{ts}\vert$} at the LHC}

\author{Ahmed~Ali}
\email{ahmed.ali@desy.de}
\affiliation{Deutsches Elektronen-Synchrotron DESY, D-22607 Hamburg, Germany}

\author{Fernando Barreiro}\email{fernando.barreiro@uam.es}
\author{Theodota Lagouri}\email{theodota.lagouri@cern.ch}
\affiliation{Universidad Autonoma de Madrid (UAM),
Facultad de Ciencias C-XI, Departamento de Fisica,
Cantoblanco, Madrid 28049, SPAIN }


\begin{abstract}
We study the prospects of measuring the CKM matrix element $\vert V_{ts}\vert$
at the LHC with the top quarks produced in the processes $p p \to t\bar{t}X$
and $p p \to t/\bar{t} X$, and the subsequent decays
$t \to W^+s$ and  $\bar{t} \to W^- \bar{s}$. To reduce the
jet activity in top quark decays, we insist on tagging the $W^\pm$ leptonically,
$W^\pm \to \ell^\pm \nu_\ell$ ($\ell =e, \mu, \tau$), and analyse the anticipated
jet profiles in the signal process $t \to W s$ and the dominant background
from the decay $t \to W b$. To that end, we analyse the $V0$ ($K^0$ and $\Lambda$)
distributions in the $s$- and $b$-quark jets concentrating on the energy and transverse
momentum distributions of these particles. The $V0$s emanating from the $t \to W b$
branch have displaced decay vertexes from the interaction point due to the weak decays
$b \to c \to s$ and  the $b$-quark jets are rich in charged leptons. Hence, the absence
of secondary vertexes and of the energetic charged leptons in the jet provide additional
($b$-jet vs. $s$-jet) discrimination in top quark decays. These distributions are used to
train a boosted decision tree (BDT), a technique used successfully in measuring the
CKM matrix element $\vert V_{tb}\vert$ in single top production at the Tevatron.
Using the BDT classifier, and a variant of it called BDTD, which makes use of
decorrelated variables, we calculate the BDT(D)-response functions corresponding to
the signal ($t \to W s$)  and background ($t \to W b$). Detailed simulations undertaken
by us with the Monte Carlo generator PYTHIA are used to estimate the background
rejection versus signal efficiency for three representative LHC energies
$\sqrt{s}=7$ TeV, 10 TeV and 14 TeV, of which only the analysis for the $\sqrt{s}=14$
TeV case is shown in detail. We argue that a benchmark with 10\% accuracy for the signal
($t \to W s $) at a background ($t \to W  b$ ) rejection by a factor $10^3$
(required due to the anticipated value of the ratio
 $\vert V_{ts}\vert^2/\vert V_{tb} \vert^2 \simeq 1.6 \times 10^{-3}$)
can be achieved at the LHC@14 TeV with an integrated luminosity of 10 fb$^{-1}$. 
\end{abstract}

\maketitle

\section{Introduction}
It is now fifteen years that the top quark was discovered in proton-antiproton
collisions at the Tevatron~\cite{Abe:1995hr,Abachi:1995iq}. Since then, a lot of
precise measurements have been undertaken at the two Fermilab experiments, CDF and D0.
Among the highlights are the  measurements of the top quark mass,
currently having  an accuracy of about 0.75\%, the $t\bar{t}$ production cross section
with about 9\% accuracy~\cite{:2009ec}, and the observation of the electroweak single
top production~\cite{Abazov:2009ii,Aaltonen:2009jj}. Of these, the single top (or anti-top)
production cross section depends on the charged current couplings $tqW$, where $q=d,s,b$,
which in the standard model (SM) are governed by the Cabibbo-Kobayashi-Maskawa (CKM)
quark mixing matrix $V_{\rm CKM}$~\cite{Cabibbo:1963yz,Kobayashi:1973fv}:
\vspace*{-0.5cm}
\begin{eqnarray*}
V_{\rm CKM} \equiv \left(\begin{matrix}
 V_{ud} & V_{us} &V_{ub} \cr
 V_{cd} & V_{cs} &V_{cb} \cr
 V_{td} & V_{ts} &V_{tb}\end{matrix}
\right)~.
\end{eqnarray*}
In the Wolfenstein Parametrisation~\cite{Wolfenstein:1983yz}, this matrix is expressed as
\begin{eqnarray*}
V_{\rm CKM} &\simeq&
  \left(\begin{matrix}
 1-{1\over 2}\lambda^2 & \lambda
 & A\lambda^3 \left( \rho - i\eta \right) \cr
 -\lambda ( 1 + i A^2 \lambda^4 \eta )
& 1-{1\over 2}\lambda^2 & A\lambda^2 \cr
 A\lambda^3\left(1 - \rho - i \eta\right) & -A\lambda^2\left(1+i \lambda^2 \eta\right) 
& 1 
\cr\end{matrix}
\right)~,
\label{eq:wolfenstein}
\end{eqnarray*}
where $A$, $\lambda$, $\rho$ and $\eta$ are the Wolfenstein parameters.

 The cross
 section $\sigma (p \bar{p} \to
t/\bar{t} X)$ has provided the first direct measurement of the dominant CKM-matrix element 
$\vert V_{tb}\vert$. In this analysis, it is assumed that the CKM matrix elements
 $\vert V_{td}\vert$ and $\vert V_{ts}\vert$ are much smaller than $\vert V_{tb}\vert$,
but no assumption is made about the unitarity of the $3\times 3$ CKM matrix. To obtain
$\vert V_{tb}\vert^2$, the measured cross section for an suumed top quark mass is divided by
 the theoretical cross section for $\vert V_{tb}\vert=1$. Following this procedure,  
 the CDF measurements yield $\vert V_{tb}\vert=0.91 \pm 0.11 ({\rm stat + syst }) \pm 0.07 ({\rm theory}$),
which in turn gives $\vert V_{tb}\vert> 0.71$ at 95\% C.L.~\cite{Aaltonen:2009jj}. The 
corresponding limit from D0 is $\vert V_{tb}\vert> 0.78$ at
 95\% C.L.~\cite{Abazov:2009ii}. The combined CDF and D0 analysis assumes 170 GeV as the
 top quark mass and yields $\vert V_{tb}\vert=0.91 \pm 0.08$ with $\vert V_{tb}\vert> 0.79$ at
 95\% C.L. using $\sigma (p\bar{p} \to t/\bar{t} +X) =3.14$ pb~\cite{Harris:2002md}.
 A theoretical value
$\sigma (p\bar{p} \to t/\bar{t} +X) =3.46$ pb~\cite{Kidonakis:2006bu}
 as input yields $\vert V_{tb} \vert=0.88 \pm 0.07$ with
 $\vert V_{tb}\vert> 0.77$ at 95\% C.L.~\cite{Group:2009qk}.
 There also exist limits on this matrix element obtained
from the decays of the top quarks by tagging the b-quark jet in the final state.
 Defining the ratio
 $R_tequiv \frac{{\cal B}(t \to bW)}
{{\cal B}(t \to dW) + {\cal B}(t \to sW) + {\cal B}(t \to bW)}=\vert V_{tb}\vert^2$, where
use has been made of the CKM unitarity in the second equality, CDF and D0 measurements
yield $\vert V_{tb}\vert > 0.78$~\cite{Acosta:2005hr} and $\vert V_{tb} \vert > 0.89$
~\cite{Abazov:2008yn}, respectively. 

The above determination of the matrix element $\vert V_{tb}\vert$, obtained from the direct
single top production and the $b$-tagged decays of the top quark, can be compared with the indirect 
determination of the same based on a number of loop-induced processes in which top quark participates
as a virtual state, such as the $B^0$- $\overline{B^0}$ and $B_s^0$- $\overline{B_s^0}$
mixings, the radiative decay $B \to X_s \gamma$ and the CP-violation parameter
$\epsilon_K$ in the Kaon sector. Overall fits of the CKM unitarity yield, comparatively speaking,
 an infinitely more accurate value $\vert V_{tb}\vert=0.999133(44)$~\cite{Amsler:2008zzb}.
This precision, in all likelihood, will not be matched by the {\it direct determination} of
 $\vert V_{tb}\vert$, as  experiments at the LHC are expected to reach an accuracy of 
a few per cent on this quantity. Neverthelss, a determination of $\vert V_{tb}\vert$ with such an accuracy will
be very valuable to constrain beyond-the-SM physics models. A good case in point is a model with
four generations, in which $\vert V_{tb}\vert$ can be as low as 0.93~\cite{Eberhardt:2010bm}.
   
We go a step further and
 explore in this paper the prospects of measuring the matrix element $\vert V_{ts}\vert$ 
at the LHC. In the Wolfenstein parametrisation~\cite{Wolfenstein:1983yz}, this matrix element
is given as $\vert V_{ts}\vert=A\lambda^2 +O(\lambda^4)$. The best-fit values from the unitarity fits
are: $A=0.814$,  $\lambda=0.2257$, yielding
 $\vert V_{ts} \vert=0.0407 \pm 0.001$~\cite{Amsler:2008zzb}. The smallest 
matrix element in the third row of the CKM matrix is $V_{td}$,  and its value from the
 CKM unitarity fit
is posted as $\vert V_{td}\vert = A\lambda^3\sqrt{(1-\rho)^2 + \eta^2}=
(8.74 ^{+0.26}_{-0.37})\times 10^{-3}$. Direct determination of these
matrix elements will require a good tagging of the $t \to s$ transition (for $\vert V_{ts}\vert$) and
$t \to d$ transition (for $\vert V_{td}\vert$) in the top quark decays,
and a very large top quark statistics, which will be available only at the LHC in the foreseeable
future from the processes $pp \to t\bar{t}+X$ and $pp \to t/\bar{t} +X$. Just as for the direct
measurement of $\vert V_{tb}\vert$, there is also a lot of interest in the direct measurements of
$V_{ts}$ and $V_{td}$, as the absolute values of these CKM matrix elements can be modified by approximately
a factor 2 from their SM values quoted above, taking the example of a four-generation extension of
the SM~\cite{Eberhardt:2010bm}.  
Lacking a good tagging for the $t \to d$ transition, and also because of the small size of the
CKM-matrix element, $\vert V_{td}\vert=O(10^{-2})$, we concentrate here on the direct measurements of
 $\vert V_{ts}\vert $ at the LHC.

In order to be able to measure $\vert V_{ts}\vert$ directly, one has to develop efficient discriminants
 to suppress the dominant decay $t \to W\; b$.
As the first step, we propose to tag only those events in which the
$W^\pm$ decay leptonically to reduce the jet activity in top quark decays.
 The emerging $s$-quark from the top quark decay $t \to W \; s$, and the collinear gluons which
are present in the fragmentation process anyway, will form a hadron jet.
  We suggest tagging on the $V0$s ($K^0$s and
 $\Lambda$s) in this jet, and measure their energy and transverse momentum distributions.
 Energetic $V0$s  are also present in the $b$-quark
jets initiated by the decay $t \to W\; b$ and the subsequent weak decays $b \to c \to s$.
 However, in this case, the $V0$s will be softer, will have displaced vertexes (from the interaction point)
 and they will be often accompanied with energetic charged leptons due to the decays
 $b \to \ell^\pm X$. Absence of a secondary vertex and paucity 
 of the energetic charged leptons in the jet provide a strong discrimination on the decays
 $t \to W b$ without essentially compromising the decays $t \to W\; s$.
 Thus,  the  scaled energy
and transverse momentum distributions of the $K^0$s, $\Lambda$s and $\ell^\pm$s, and
 the secondary decay vertex distributions ($dN/dr$) are the quantities 
of principal interest. Here $r$ is the distance traversed in the transverse plane, i.e. the
plane perpendicular to the beam axis or $r-\phi$ plane,
by the $b$-quark before decaying, smeared with a Gaussian resolution to take
into account realistic experimental conditions.

We have assumed two representative r.m.s. values ($\sigma$(vertex) =1 mm and 2 mm) for the Gaussian, where  2 mm is more realistic. Experimentally, b-tagging algorithms are based on measurements of the impact parameter from the $B$ meson charged tracks. The power separation between b- and light-jets using this so called 2D-method is similar to the one we used in our analysis with 2 mm resolution. We also show results with 1 mm resolution in order to
illustrate how important it will be for the Super LHC (SLHC) to improve on the b-tagging efficiency.
 These distributions are calculated for the processes $pp \to t \bar{t}X$ and
$pp \to t/\bar{t}X$, for the signal ($t \to W\; s $) and background ($t \to W\; b $). 

Having generated these distributions, characterising the signal $t \to W\; s$
and the background $t \to W\; b$ events, we use a technique called the
Boosted Decision Tree (BDT) -- a classification model used widely in data mining~\cite{Han:2006} 
-- to develop an identifier optimised for the $t \to W\;s$ decays. In our calculation, we use both
 BDT and a variant of it called BDTD (here D stands for decorrelated), where possible correlations
in the input variables are removed by a proper rotation obtained from the decomposition 
of the square root of the covariance matrix, to
discriminate the signal events from the large backgrounds.  We recall that
this technique has been successfully used to establish the single top quark production in $p \bar{p}$
collisions at the Tevatron~\cite{Abazov:2009ii,Aaltonen:2009jj} (see~\cite{Liu:2009zz} for details).
 Briefly, the generated input is used for the purpose of training and testing the samples.
 We provide the input in terms of the variables discussed earlier for the signal
($ t \to Ws$) and the background ($ t \to Wb$), obtained with the help of a Monte Carlo
generator. This information is used to develop the splitting criteria to determine the
best partitions of the data into signal and background to build up a decision tree (DT). 
The separation algorithm
used in splitting the group of events in building up DT plays an important role in the performance.
 The software 
called the Toolkit for Multivariate Data Analysis in ROOT (TMVA)~\cite{Hocker:2007ht} is used for the
 BDT(D) responses in our analysis.
 Detailed simulations presented here are done using
 PYTHIA~\cite{Sjostrand:2007gs} to model the production processes, gluon radiation, fragmentation and
 decay chains, and the underlying events. We calculate the signal ($t \to W\; s  $) efficiencies
for two cases called $bb/bs$ and $bs/ss$ (defined in section II) for an assumed (Gausian) vertex
smearing with an r.m.s. value of 2mm and 1mm. Concentrating on the $bb/bs$ case, when only one of
the top (or antitop) quark decays via $t \to s W^+$, compared to the case when both the top and
antitop decay via the dominant transition $t \to bW^+$, these efficiencies lie typically between
5\% (for the 2mm smearing) and 20\% (for the 1mm case) for a background ($ t \to W\; b$) rejection by
 a factor $10^3$ (see Table I).

Note that this level of background rejection is 
necessary due to the anticipated value of
the ratio $\vert V_{ts}\vert^2/\vert V_{tb} \vert^2 \simeq 1.6 \times 10^{-3}$.
The required integrated LHC luminosity to determine $\vert V_{ts}\vert$ {\it directly} is estimated
as 10 fb$^{-1}$ at 14 TeV. Numerical analysis reported here
is carried out for three representative  LHC energies:  $\sqrt{s}=7$ TeV,  10 TeV and 14 TeV, but
we present the detailed results only for $\sqrt{s}=14$ TeV as the distributions for
 $\sqrt{s}=7$ TeV and 10 TeV are similar to the 14 TeV case.

In section 2, we study the process 
$pp \to t\bar{t}X$, reviewing first the production cross sections at the LHC energies.
The energy-momentum profiles of the signal ($t \to W\; s$) 
and background ($ t \to Wb$) events 
produced in the $t\bar{t}$  pair production process and the subsequent decays
 $p p \to t (\to W^+b)\; \bar{t}(\to W^-\bar{b})X$,  $p p \to t (\to W^+b)\; \bar{t}(\to W^-\bar{s})X$,
 $p p \to t (\to W^+s)\; \bar{t}(\to W^-\bar{b})X$   and
 $p p \to t (\to W^+s)\; \bar{t}(\to W^-\bar{s})X$ are worked out.
 The last of these has a very small branching ratio and its measurement would require
a huge LHC luminosity (we included this case for the sake of completeness).
  Tagging efficiencies for $pp \to t\bar{t}X$ calculated with the BDT(D) classifier are shown
in Table 1. Numerical results in this table are presented as $(bb/bs)$ and $(bs/ss)$, corresponding to
 the cases when only one of the $t$ (or $\bar{t}$) decays via $t \to W^+\; s$
 (or $\bar{t} \to W^-\;\bar{s}$) and  when both the $t$ and $\bar{t}$ decay via
$t \to W^+\; s$ and $\bar{t} \to W^-\;\bar{s}$, respectively. 

Section 3 is a repeat of the above analysis for the single top production process $pp \to t/\bar{t}X$ at
the LHC. The end-product of this analysis chain is again the  background rejection vs. the signal
 efficiency based on the BDT(D) response functions. The tagging efficiencies, calculated for
 $\sqrt{s}=7$, 10, and 14TeV with the BDTD classifier are presented in Table 2.
 Section 4 briefly summarises our results and outlook. 

\section{Analysis of the process $ pp \to t \bar{t}X$ and the subsequent decays $t \to Wb, Ws$}
Theoretical predictions of the top quark production at the LHC have been
obtained by including up to the next-to-next-to-leading order (NNLO) corrections in the
strong coupling
 constant~\cite{Bonciani:1998vc,Cacciari:2008zb,Kidonakis:2008mu,Moch:2008qy}.
They have been updated taking into account  modern parton distribution
functions (PDFs)~\cite{Martin:2007bv,Nadolsky:2008zw}.  A typical estimate is:
$\sigma(pp \to t\bar{t}X)= 874^{+14}_{-33}$pb for
$m_t=173$ GeV and $\sqrt{s}=14$ TeV~\cite{Langenfeld:2009tc}, where the errors
reflect the combined uncertainties in the factorisation and normalisation scales and in
the parton distribution functions (PDF). Other independent NNLO calculations 
yield similar cross section, though the error budgeting is somewhat different.
Kidonakis and Vogt~\cite{Kidonakis:2008mu} put the cross section 
 $\sigma(pp \to t\bar{t}X)=894 \pm 4({\rm kinematics}) ^{+68}_{-44}({\rm scale})
^{+29}_{-31} ({\rm PDF})$ for the same values of $m_t$ and $\sqrt{s}$, using the
CTEQ6.6M PDFs~\cite{Nadolsky:2008zw}, and 
$\sigma(pp \to t\bar{t}X)=943 \pm 4({\rm kinematics}) ^{+77}_{-49}({\rm scale})
\pm 12 ({\rm PDF})$, using the MRST 2006 PDFs~\cite{Martin:2007bv}.
Compared to the $t\bar{t}$ production cross section at the Tevatron
($\sqrt{s}=1.96$ TeV),
 $\sigma(p\bar{p} \to t\bar{t}X)=7.34^{+0.23}_{-0.38}$ pb~\cite{Langenfeld:2009tc},  
one expects a rise in the $t\bar{t}$ cross section by more than two orders of
magnitude between the Tevatron and the LHC@14 TeV. The cross sections at the lower
LHC energies, 7 and 10 TeV, have also been
 calculated~\cite{Langenfeld:2009tc,Kidonakis:2008mu}, with
$\sigma (pp \to t\bar{t}X) \simeq 400$ pb at 10 TeV and about half that number
at 7 TeV. Thus, for the top quark physics, the dividends in going from 7 to 14 TeV
are higher by a good factor 4. 

For the numerical results shown here we have used the PYTHIA Monte Carlo~\cite{Sjostrand:2007gs} to generate $10^6$ events
for the process $ p p \to t \bar{t}X$, followed by the decay chains $t \to W^+b,\; W^+s$ and
$\bar{t} \to W^-\bar{b}, \; W^- \bar{s}$. As stated in the introduction, the $W^\pm$
are forced to decay only leptonically
$W^\pm \to \ell^\pm \nu_\ell$ ($\ell =e, \mu, \tau $) to reduce the jet activity from the non-leptonic
decays of the $W^\pm$. This corresponds to an integrated luminosity of 10 fb$^{-1}$ at 14 TeV. For an
estimated efficiency of 5\% at a $10^{-3}$ background rejection, and $\vert V_{ts}\vert^2 \simeq
1.7 \times 10^{-3}$, as in the SM, this means that we
expect $0.05 \times 2\times 1.7 \times 10^{-3}\times 10^6= 170$ signal events with a background of
 $10^{-3}\times 10^6=10^3$ events, giving a significance of
$170/\sqrt{1000}$ i.e. more than $5\sigma$. 

  We then concentrate on the
$V0$ production, which for the experimental conditions at the two main detectors ATLAS~\cite{Aad:2008}
 and the CMS~\cite{Adolphi:2008} implies
$V0=K_S^0$ or $V0=\Lambda$, as the long-lived $K_L^0$ will decay mostly out of the detectors. However,
both $K_S^0$ and $\Lambda$ can be detected by ATLAS and CMS and their energy and momentum measured with reasonably
good precision. In the present analysis, we reconstruct $V0$s and soft leptons in the rapidity range
 $|\eta| \leq 2.5$~\cite{Aad:2008}.
In addition, we require the $V0$'s decay radius to lie in the range 20 to 600 mm. These acceptance cuts are
acceptable for both multipurpose detectors mentioned above, and they will be used in the analysis described in this
and the next section.
 
\begin{figure}
\center{
\includegraphics[width=0.95\textwidth]{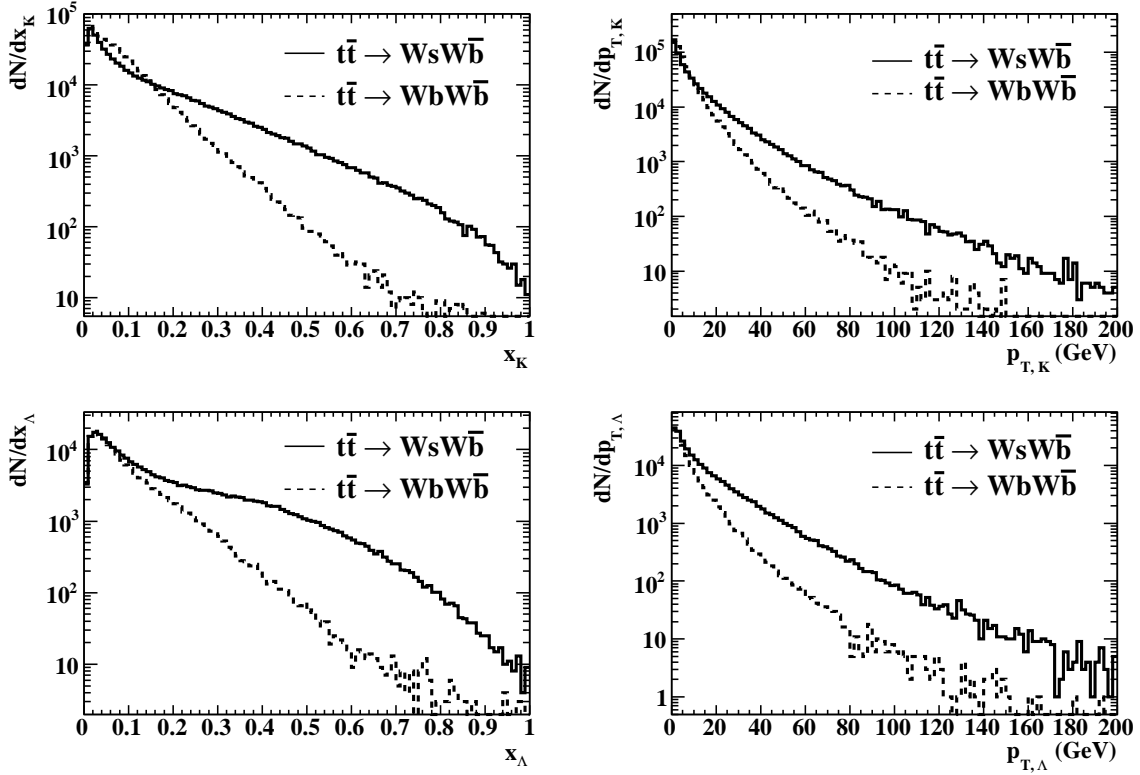}}
\caption{\label{Fig-1}
$pp \to t\bar{t}X$ at $\sqrt{s}=14$ TeV. Upper left frame: scaled-$K^0$-energy distributions $dN/dx_K$  
from  $t \to W\; s (\to K^0 X)$ (solid histogram) and $t \to W\; b(\to K^0 X)$ (dashed histogram).
 Upper right frame: Transverse momentum
distributions of the $K^0$s measured w.r.t. beam axis $dN/dp_{T_K}$ (in GeV)
 in the same production and decay processes as
in the left frame. Lower frames show the distributions $dN/dx_\Lambda$ and  $dN/dp_{T_\Lambda}$ (in GeV)
for $t \to W\; s (\to \Lambda X)$ (solid histogram) and $t \to W\;b (\to \Lambda X)$ (dashed histogram).
} 
\end{figure}

 We will show the distributions for $\sqrt{s}=14$ TeV, the designed LHC center-of-mass energy.
 The $K^0$-energy distribution is shown in the left-hand frame in Fig~\ref{Fig-1}
plotted  as a function of the scaled energy  $X_K= E_K/E_{\rm jet}$. For this study, the jet energy
 is set equal to the quark energy produced in the decay $t \to Wb, Ws$. In a realistic simulation of the
 experimental measurements, one
would require a functional definition of the jet, for example using an angular cone, which will then
define the jet energy $E_{\rm jet}$, and hence $x_K$. The transverse momentum of the $K^0$s,
$p_T(K^0)$ (in GeV), is shown in the
right hand frame in  Fig~\ref{Fig-1}. In both of these frames, the solid histograms correspond to the
decay $t \to W\; s$ and the dashed ones to the decay $t \to W\; b$. As expected, the decay chain $t \to W s (\to K_S^0)$
has a much stiffer distribution both in $X_K$ and $p_T(K^0)$, as the $K^0$'s descending from the decay
chain $t \to W\; b (\to c \to s)$ are rapidly degraded in these variables due to the subsequent weak decays. The
corresponding distributions for the $\Lambda$s are shown in the lower two frames in  Fig.~\ref{Fig-1}.
 They are qualitatively very similar to those of the $K^0$s. 

\begin{figure}
\center{
\includegraphics[width=0.70\textwidth]{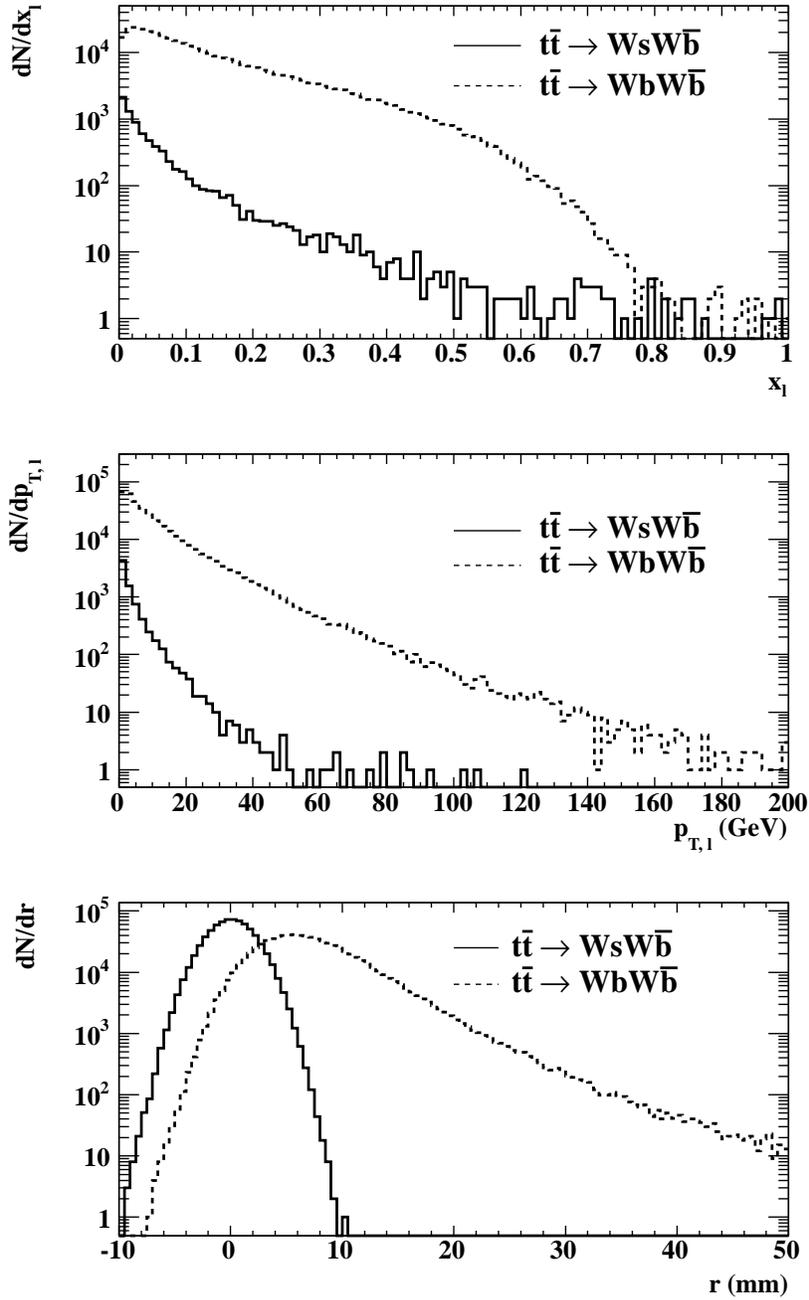}}
\caption{\label{Fig-2}
$pp \to t\bar{t}X$ at $\sqrt{s}=14$ TeV. Upper frame:  scaled-$\ell^\pm$-energy distributions, $dN/dx_\ell$, 
 from $t \to W\; s (\to \ell^\pm X)$ (solid histogram) and $t \to W\; b(\to \ell^\pm X)$ (dashed histogram).
Middle frame: Transverse momentum
distributions of the $\ell^\pm$s measured w.r.t. beam axis, $dN/dp_{T_\ell}$ (in GeV),
 in the same production and decay processes as
in the upper  frame.
Lower frame: Secondary decay vertex distributions in the variable $r$ (in millimetres) for the two decay chains
$t \to W\; s$ (solid histogram) and $t \to W\; b$ (dashed histogram), obtained by smearing the
decay length with a Gaussian having an r.m.s. value of 2 mm.  
 } 
\end{figure}

 We now show the distributions in the charged lepton energy from the
decays $t \to b \to \ell^\pm X$ and $t \to s \to \ell^\pm X$ in Fig.~\ref{Fig-2}, showing the scaled
lepton energy in the variable $X_\ell=E_\ell/E_{\rm jet}$ (upper  frame) and in $p_T^\ell$, the transverse
momentum of the charged leptons (middle frame). This distribution quantifies the richness of the
 $b$-jets in
charged leptons and the stiff character of the energy/transverse momentum distributions due to the
 weak decays,
as compared to the leptons from $s \to \ell^\pm X$, which are all soft and coming from the leptonic decays
of the various resonances produced in the fragmentation of the $s$-quark. Absence of energetic charged leptons
in the $s$-quark jet in the decay $t \to W\; s$ is a powerful tool in reducing the background from the otherwise
much more prolific process $t \to W\; b$. The final set of distributions from our Monte Carlo simulation is
the secondary decay vertex distribution (lower frame), smeared with a Gaussian distribution with a r.m.s. of 2
millimetres, shown in terms of a variable called $r$ (measured in millimetres).
The decay length for the $t \to W\;b$ case is calculated as $\gamma c \tau_b$, where $\gamma$ is the Lorentz factor,
and  $c\tau_b=0.45$ mm, corresponding 
to an average $b$-quark lifetime taken as
$\tau_b=1.5$  ps from the PDG~\cite{Amsler:2008zzb}. This distribution, which reflects the long lifetime of the $b$-quark (respectively of
the $B$ and $\Lambda_b$ hadrons), as opposed to the lack of a secondary vertex from 
the $s$-quark fragmentation process, is also a very powerful discriminant of $t \to W\; b$ vs. $t \to W\; s$
decays.


\begin{figure}[htb]
\centerline{
\hskip 0.cm
\epsfysize 6.0 cm
\epsffile{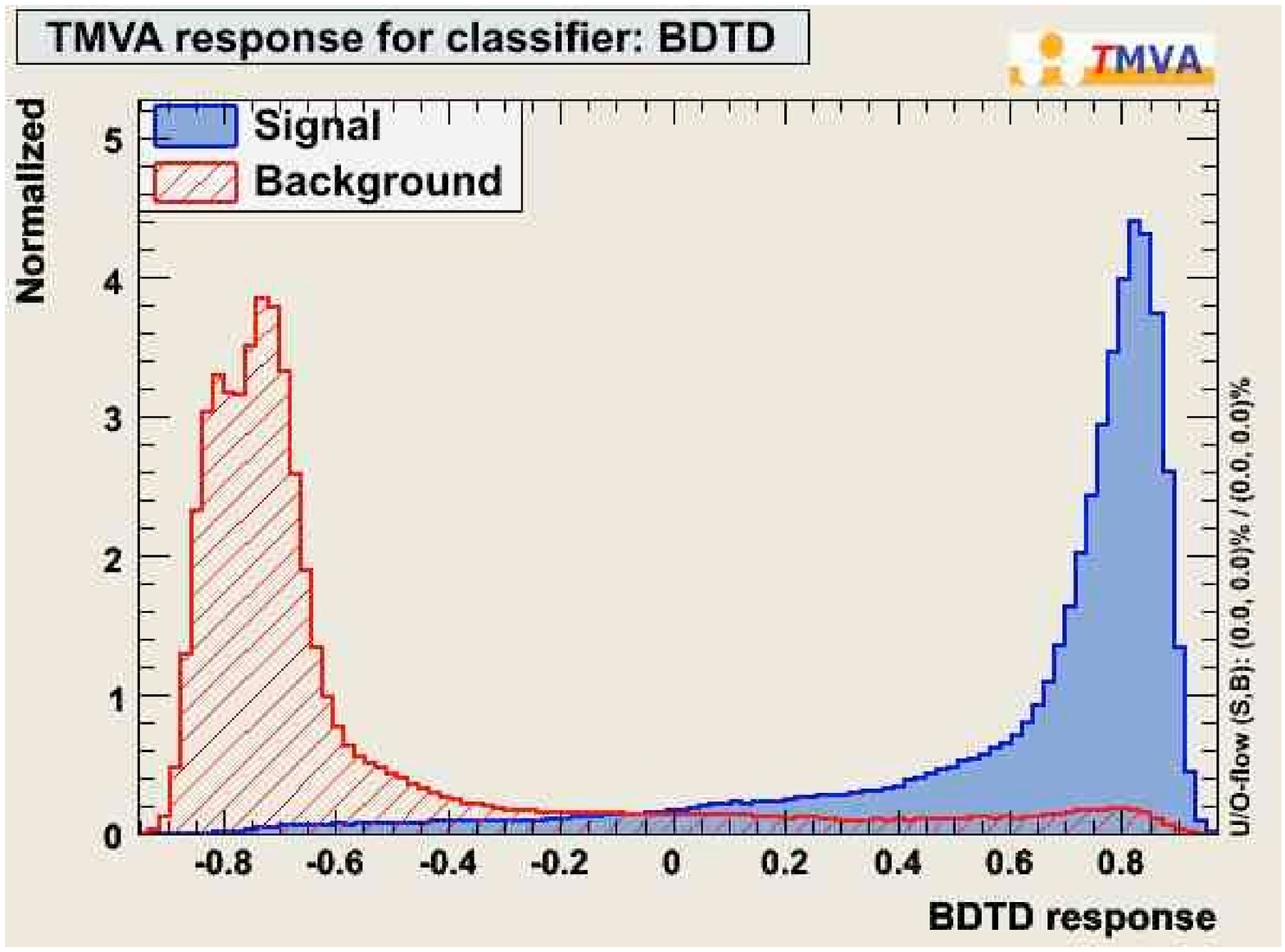}
\hskip 0.0cm
\epsfysize 6.0 cm
\epsffile{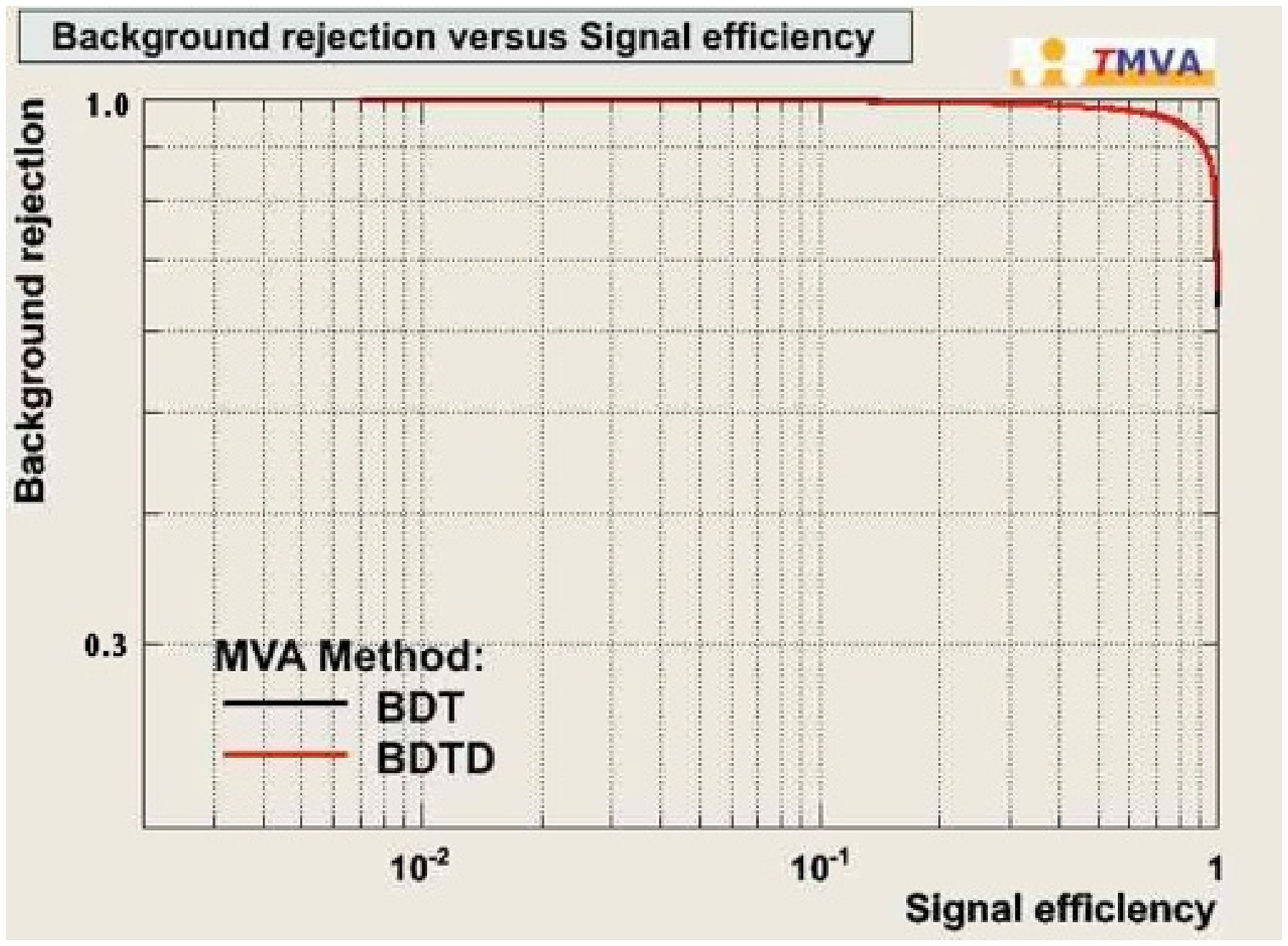}}
\vspace*{-0.2cm}
\caption{\label{Fig-3}
$pp \to t\bar{t}X$ at $\sqrt{s}=14$ TeV. Left frame: The normalised BDTD response, calculated by using the TMVA (see text). The signal (dark shaded) from the decay $t \to W\; s$ and the background
(light shaded with dotted lines) from the decay $t \to W\; b$ are clearly separated in this variable.
 Right frame: Background rejection vs. signal efficiency
calculated from the BDT(D) response. The result using the BDT classifier is very similar
and hardly distinguishable from the one obtained with the BDTD response.} 
\end{figure}

Having generated these distributions, characterising the signal $t \to W\; s$
and background $t \to W\; b$ events in the process $pp \to t \bar{t} X$ at the LHC, we use the
BDT and BDTD classifiers, discussed in the introduction. 
In Fig.~\ref{Fig-3} (left frame), we show the BDTD response functions, showing that a clear separation
between the signal ($ t \to W\; s$) and background ($ t \to W\; b$) events has been achieved.
 The background rejection vs. signal efficiency for
 the $pp \to t\bar{t}$
events is shown in Fig.~\ref{Fig-3} (right frame) for both the BDT and
BDTD classifiers, which give very similar results. The evaluation results ranked by the best signal efficiency and purity
are shown numerically in Table 1. The entries in this table
show that a background rejection of $10^{3}$ can be achieved at a signal efficiency of about $5\%$ 
to reach the SM-sensitivity of the CKM matrix element $\vert V_{ts}\vert$. The statistical uncertainty in
this efficiency is $0.026\%$. It goes up to $0.067\%$ for efficiency values at the upper end.

The distributions at $\sqrt{s}=7$ and 10 TeV are very similar to the corresponding ones shown in
Fig.~\ref{Fig-1} for 14 TeV. Hence, the characteristic differences that we have
shown at $\sqrt{s}=14$ TeV emanating from the top quark decays $t \to W b$ and $t \to W s$ in the
$V0$ and charged lepton energy- and transverse momentum  spectra are also present at the
lower energies.

Based on the above analysis we have calculated the tagging efficiencies for the decay $t \to W\;s$
 (signal)
for an acceptance of 0.1\% for the decay $t \to W b$ (background). The acceptance level is motivated by
the anticipated value of the ratio of the $t \to W s$ and $t \to W b$ decay rates, which in the SM is
$O(10^{-3})$.  The tagging efficiencies for
the three centre-of-mass energies at the LHC (7, 10 and 14 TeV) are given in Table 1 for two different
vertex smearing (1 mm and 2 mm), assuming a Gaussian distribution. The entries shown as $bb/bs$
 correspond to the comparison for top pair production process $pp \to t \bar{t}X$
with both the $t$ and $\bar{t}$ decaying via the dominant process $t \to W^+b$
and  $\bar{t} \to W^- \bar{b}$, respectively, and in which only one of the
 $t$ or $\bar{t}$ quarks
decays via $t \to W^+ s$ or $\bar{t} \to W^- \bar{s}$,  and the other decays via $t \to W^+ b$
or $\bar{t} \to W^- \bar{b}$ (signal events). The entries marked as
$bs/ss$ correspond to the cases where either the $t$ or $\bar{t}$ quarks decays via $t \to W^+ s$
or $\bar{t} \to W^- \bar{s}$ and
both $t$ and $\bar{t}$ quarks decay via $t \to W^+s$ and $\bar{t} \to W^- \bar{s}$. The branching ratio for the
 case $(t \to W^+ s))(\bar{t} \to W^- \bar{s})$ is exceedingly small, $O(10^{-6})$. 
The entries in Table 1 for this case ($bs/ss$) show that at the considerable price of
the reduced sample, one can get much better efficiencies.   
\begin{table}
\begin{center}
\caption{Tagging efficiencies (in \%) for the process $pp \to t \bar{t}X$, followed by the
decay $t\rightarrow Ws$ (signal) and $t\rightarrow Wb$( background),
 calculated for an acceptance of $0.1\%$ for the background at three LHC centre-of-mass
energies. Two Gaussian vertex smearing (having an r.m.s. values of 2 mm and 1 mm) are assumed for
calculating the displaced vertex distributions $dN/dr$. The cases $bb/bs$ and $bs/ss$ are explained in the text.}  
\vspace*{0.5cm}

\sf

\begin{tabular}{|l|l|l|l|r|}

\hline

$bb/bs$ & vertex smearing    & 7 TeV & 10 TeV & 14 TeV   \\

\hline

 &2 mm   & 5.1 & 5.6 & 5.0 \\

\hline

 &1 mm  & 20.5 & 15.4 & 15.5  \\

\hline

$bs/ss$ & vertex smearing    & 7 TeV & 10 TeV & 14 TeV   \\

\hline

&2 mm   & 13.2 & 9.6 & 12.3 \\

\hline

&1 mm  & 30.6 & 24.2 & 34.2  \\

\hline

\end{tabular}
\end{center}
\end{table}
\vspace*{0.5 cm}

\section{Analysis of the process $ pp \to t/\bar{t}X$ and the subsequent decays $t \to Wb, Ws$}
The single top (or anti-top) cross sections in hadron hadron collisions have been
calculated in the NLO
 approximation~\cite{Harris:2002md,Cao:2004ap,Heim:2009ku,Kidonakis:2006bu,Kidonakis:2007ej}.
 Recalling that there are 
three basic processes at the leading order which contribute to
 $\sigma (p\bar{p} \to t/\bar{t} X)$, namely the $t$-channel: $qb \to q^\prime t$,
the $s$-channel: $q\bar{q}^\prime \to \bar{b}t$; and the associated $tW$ production
$bg \to tW^-$, the cross sections estimated at the Tevatron are~\cite{Kidonakis:2009sv}:
 $\sigma^{t-{\rm channel}} (p\bar{p} \to t X)=\sigma(p \bar{p} \to \bar{t}X)=
1.14 \pm 0.06$ pb,
 $\sigma^{s-{\rm channel}} (p\bar{p} \to t X)=\sigma(p \bar{p} \to \bar{t}X)=
0.53 \pm 0.02$ pb, and
  $\sigma^{tW-{\rm channel}} (p\bar{p} \to t X)=\sigma(p \bar{p} \to \bar{t}X)=
0.14 \pm 0.03$ pb, putting the single top (or antitop) cross section
 $\sigma (p\bar{p} \to t X)=\sigma(p \bar{p} \to \bar{t}X) \simeq 1.8$ pb at the
Tevatron.
These cross sections have to be compared with the theoretically projected
cross sections at the LHC@14 TeV: $\sigma^{t-{\rm channel}} (p p \to t X)=149 \pm 6$ pb, 
 $\sigma^{t-{\rm channel}} (p p \to \bar{t} X)=91 \pm 4$ pb,  
$\sigma^{s-{\rm channel}} (p p \to t X)=7.7 ^{+0.6}_{-0.7}$ pb,
$\sigma^{s-{\rm channel}} (p p \to \bar{t} X)=4.3 \pm 0.2$ pb, and
$\sigma^{tW-{\rm channel}} (p p \to t X)=\sigma(p p \to \bar{t}X)=43 \pm 5$ pb.
Thus, one expects $\sigma(pp \to tX) \simeq 200$ pb and about half this number for
$\sigma(pp \to \bar{t}X)$, yielding the summed single top and antitop cross sections at
about 300 pb at the  LHC@14 TeV, also approximately two orders of magnitude larger
than those at the Tevatron. With a luminosity of 10 fb$^{-1}$, one anticipates 
$O(3 \times 10^6)$ single top (or anti-top) events, i.e. $O(10^6)$ events in the
leptonic channel. Thus, the rise in the cross sections
for a single top (or antitop) production between the Tevatron and the LHC@14 TeV is also 
very marked.

\begin{figure}
\center{
\includegraphics[width=0.95\textwidth]{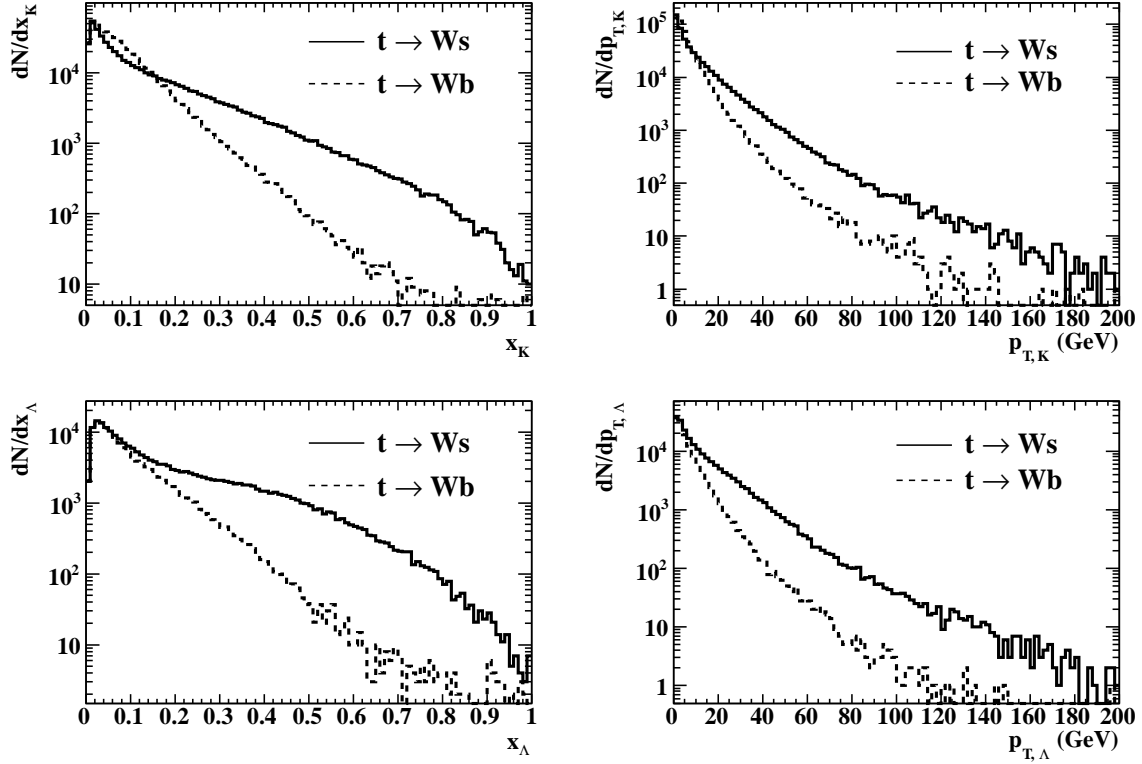}}
\caption{\label{Fig-9}
$pp \to t/\bar{t} X$ at $\sqrt{s}=14$ TeV.
Upper frames: Scaled energy distributions $dN/dx_K$ and the transverse momentum 
distribution $dN/dp_{T_K}$ from the decays  $t \to Ws (\to K^0 X)$ (solid histograms) and
$t \to Wb(\to K^0 X)$ (dashed histograms).
 Lower frames: Scaled energy distributions $dN/dx_\Lambda$ and the transverse momentum 
distribution $dN/dp_{T_\Lambda}$ from the decays  $t \to Ws (\to \Lambda X)$ (solid histograms) and
$t \to Wb(\to \Lambda X)$ (dashed histograms).
} 
\end{figure}
\begin{figure}
\center{
\includegraphics[width=0.70\textwidth]{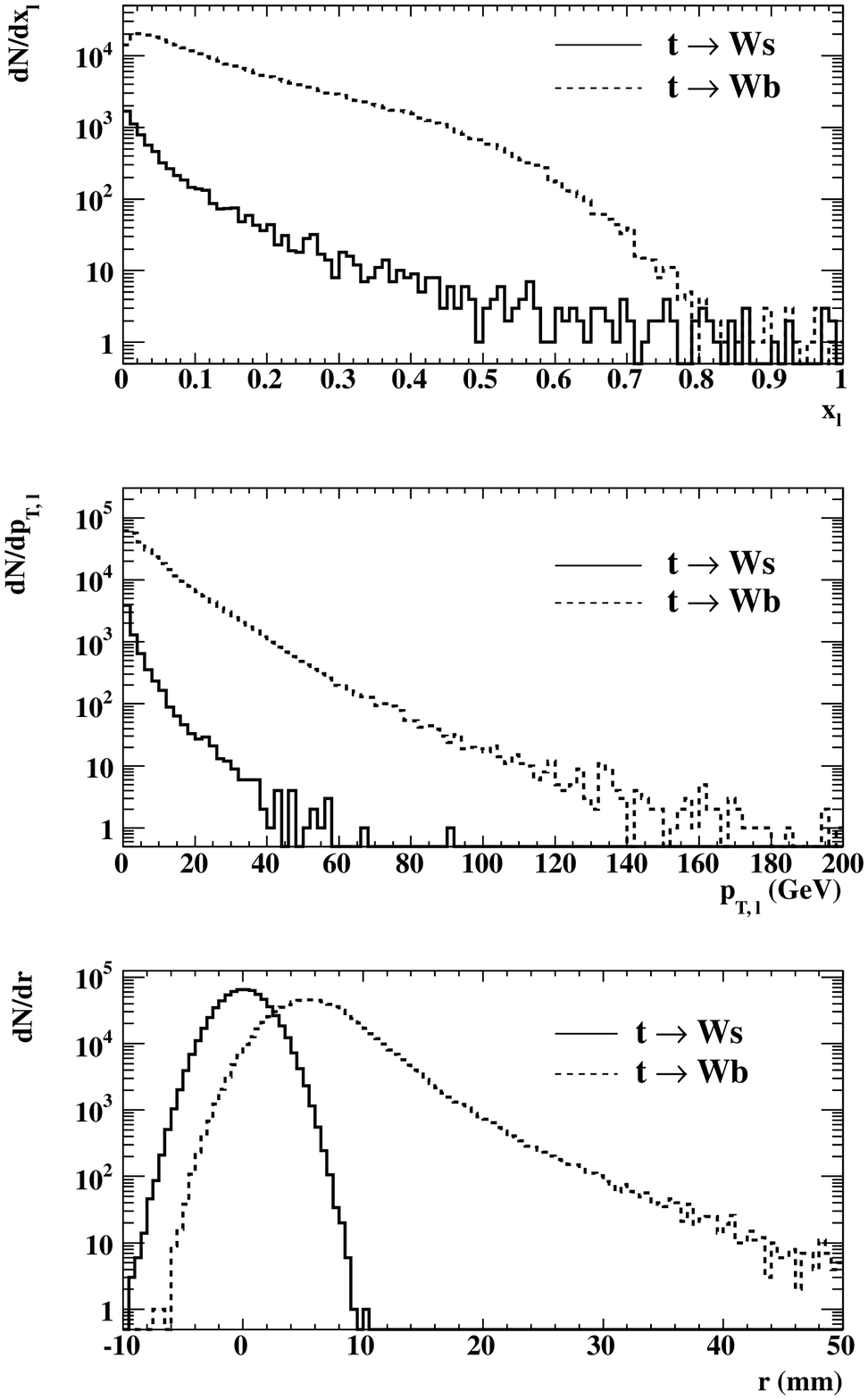}}
\vspace*{-0.2cm}
\caption{\label{Fig-10}
 $pp \to t/\bar{t} X$ at $\sqrt{s}=14$ TeV. 
Upper frame:  scaled-$\ell^\pm$-energy distributions, $dN/dx_\ell$, 
 from $t \to W\; s (\to \ell^\pm X)$ (dashed histogram) and $t \to W\; b(\to \ell^\pm X)$ (solid histogram).
Middle frame: Transverse momentum
distributions of the $\ell^\pm$s measured w.r.t. beam axis, $dN/dp_{T_\ell}$ (in GeV),
 in the same production and decay processes as
in the upper  frame.
Lower frame: Secondary decay vertex distributions in the variable $r$ (in millimetres) for the two decay chains
$t \to W\; s$ (solid histogram) and $t \to W\; b$ (dashed histogram), obtained by smearing the
decay length with a Gaussian having an r.m.s. value 2 millimetres. 
}  
\end{figure}

\begin{figure}[htb]
\vspace*{-1.0cm}
\centerline{
\hskip 0.cm
\epsfysize 6.0 cm
\epsffile{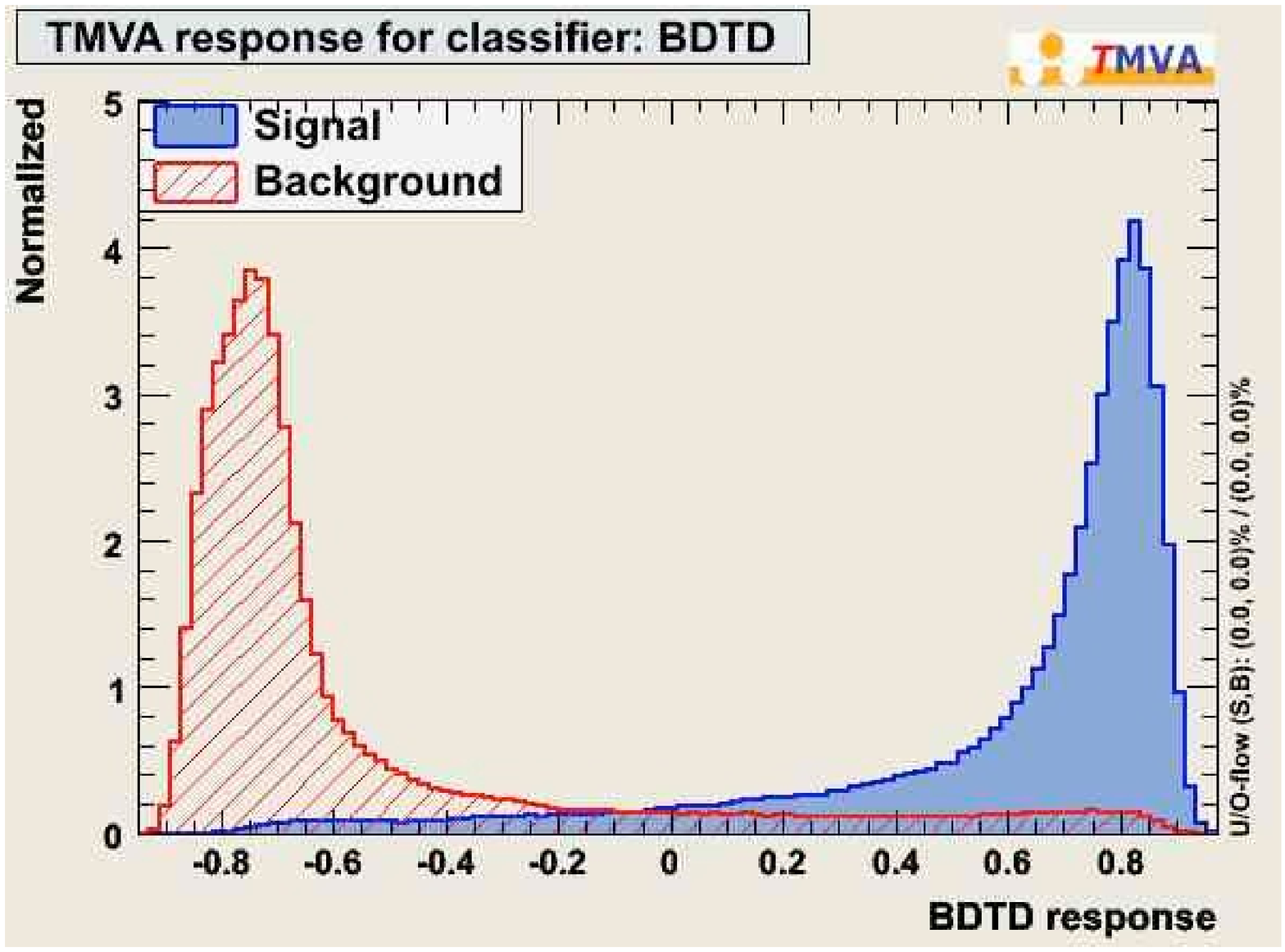}
\hskip 0.0cm
\epsfysize 6.0 cm
\epsffile{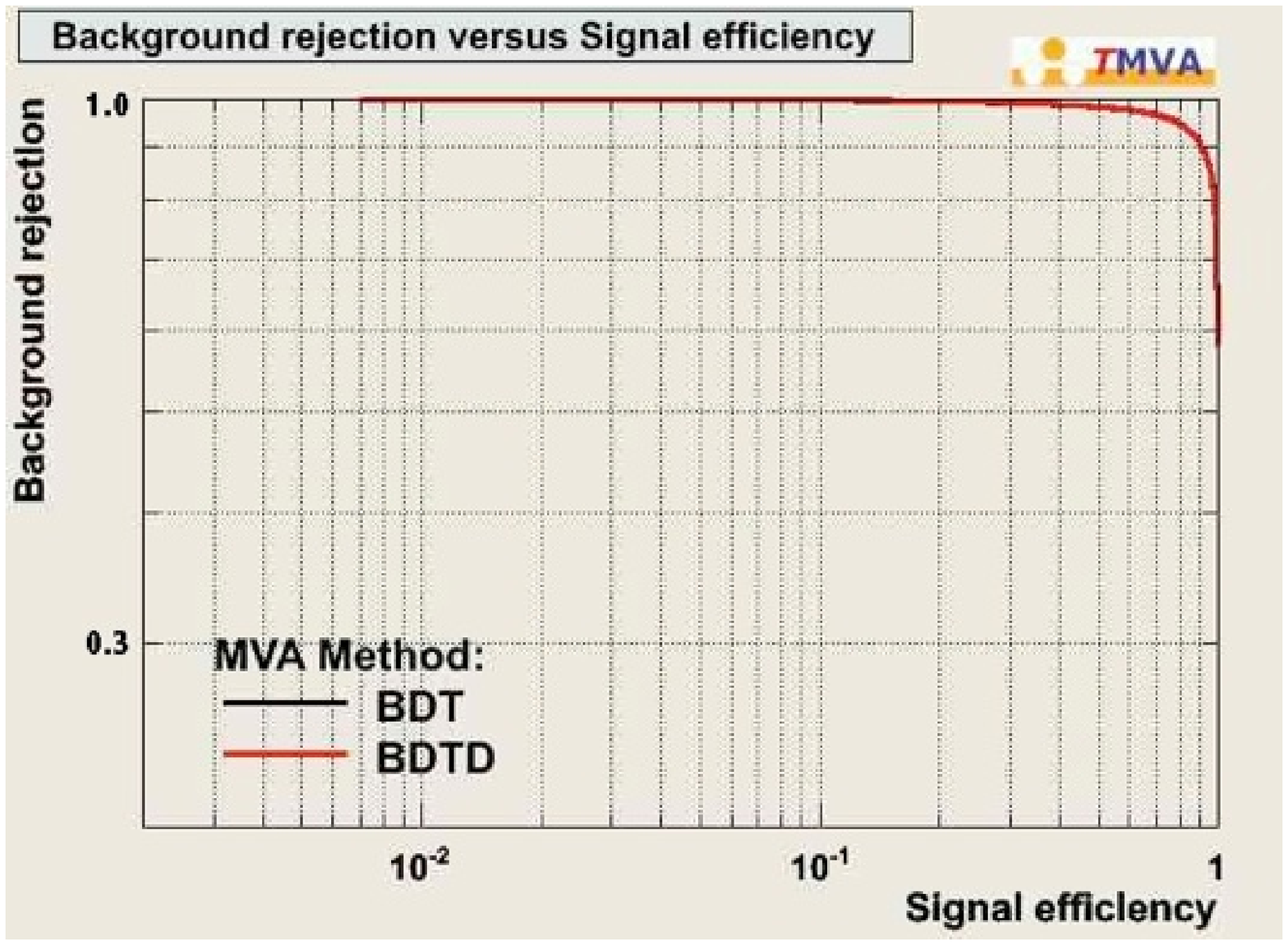}}
\vspace*{-0.2cm}
\caption{\label{Fig-11}
$pp \to t/\bar{t}X$ at $\sqrt{s}=14$ TeV. Left: The normalised BDTD response, calculated by using the TMVA (see text). The signal (dark shaded) from the decay $t \to W\; s$ and the background
(light shaded with dotted lines) from the decay $t \to W\; b$ are clearly separated in this variable. Right: Background rejection vs. signal efficiency
calculated from the BDT(D) response. The two MVA methods
yield very similar results. }
\end{figure}

In the analysis shown here, we have again resorted to the Monte Carlo generator PYTHIA, which
models so far only the $s$-channel single top production process $pp \to W \to t\bar{b}$. As discussed
above, this has the smallest (of the three channel) cross section. Moreover, the $tW$ channel process  
$bg \to tW^-$ provides a much more efficient trigger in terms of the $W^-$ accompanying the top quark.
So, the analysis presented in this section should be repeated with a more complete code, including
all three channels. However, we think that in estimating the various efficiencies, the current version
of PYTHIA is adequate.  The distributions in the scaled energy variable $X_K$ and in the transverse momentum
$p_T$ of the $K^0$s produced in the process $pp \to t/\bar{t} X$, and the subsequent decays
$t \to W\; s$ and $t \to W\; b$ are shown in Fig.~\ref{Fig-9} (upper two frames) for $\sqrt{s}=14$ TeV. The corresponding
 distributions for the $\Lambda$s
from the decays $t \to Ws (\to \Lambda X)$ and $t \to Wb (\to \Lambda X)$ are shown in the lower two frames
in Fig~\ref{Fig-9}. 
The scaled charged lepton energy distribution $X_\ell =E_\ell/E_{\rm jet}$ and the lepton
transverse momentum are shown in Fig.~\ref{Fig-10} (upper frame), showing the distributions from the 
$t \to W b (\to \ell^\pm X)$ (dashed histograms) and $t \to W s (\to \ell^\pm X)$  (solid histograms).
Transverse momentum distributions of the $\ell^\pm$s, measured w.r.t. the axis, are also shown in
this figure (middle frame). 
Finally, the secondary decay vertex distribution for the decays $t \to Wb $ (dashed histogram)
 and $t \to Ws $ (solid histogram) are shown in the lower frame in this figure. In plotting these
distributions, we have smeared them, as before,  with a Gaussian with a r.m.s. value of 2 millimetres
 and have taken into
account the finite lifetime of the $b$-quark, as stated in the case of the analysis for the process
$pp \to t\bar{t} X$. These distributions provide an excellent discrimination between the signal
$t \to W\; s$ and the dominant background $t \to W\; b$ events for the single top production process
$pp \to t/\bar{t} X$, qualitatively very much along the same lines
as discussed earlier for the $t \bar{t}$ production $p p \to t \bar{t}X$. As already stated,
this information is used to  build up a decision tree . 
In Fig.~\ref{Fig-11} (left frame), we show that the BDTD response function is very different for the signal ($ t \to Ws$)
and the background ($ t \to Wb$) events also for the single top (antitop) production process.
 The background rejection vs. signal efficiency for the $pp \to t/\bar{t} X$
events is shown in Fig.~\ref{Fig-11} (right frame).   The  results for the signal efficiency and purity
are shown numerically in Table II.  The entries in this table 
show that also in single top production process a background rejection of $10^{3}$ can be achieved at a signal efficiency of about 5\% to 7\%
to reach the SM-sensitivity of the CKM matrix element $\vert V_{ts}\vert$. Thus we would expect
$0.053 \times 1.7 \times 10^{-3}\times 10^6= 90$ signal events with a background of
 $10^{-3}\times 10^6=10^3$ events giving a significance of
$90/\sqrt{1000}$ i.e. about $3\sigma$. 

\begin{table}
\begin{center}
\caption{Tagging efficiencies (in \%) for the single top production process $pp \to t/ \bar{t}X$,
 followed by the
decay $t\rightarrow Ws$ (signal) and $t\rightarrow Wb$( background),
 calculated for an acceptance of $0.1\%$ for the background at three LHC centre-of-mass
energies. Two Gaussian vertex smearing (having an r.m.s. values 2 mm and 1 mm) are assumed for
calculating the displaced vertex distributions $dN/dr$.}
\vspace*{0.5cm}

\sf

\begin{tabular}{|l|l|l|r|}

\hline

 $\sigma_0$    & 7 TeV & 10 TeV & 14 TeV   \\

\hline

 2 mm   & 7.1 & 6.5 & 5.3 \\

\hline

 1 mm  & 21.7 & 22.4 & 19.9  \\

\hline

\hline

\end{tabular}
\end{center}
\end{table}
\vspace*{0.5 cm}
\section{Summary and Outlook}

We have presented a case here to measure the matrix element $\vert V_{ts} \vert$ from the top quark decays
$t \to W^+s$ and its charge conjugate $\bar{t} \to W^- \bar{s}$, making use of the characteristic
differences in the $b$- and $s$-jet profiles. We have concentrated on the V0 $(K^0$ and $\Lambda$)
energy-momentum profiles emanating from the signal ($ t \to Ws$) and the dominant background
($t \to b W $). This information is combined with the secondary vertex distributions, anticipated
from the decays ($b \to c \to s$), and the absence of energetic charged leptons in $s$-quark jets.
An important parameter is the vertex resolution, for which we have used two values, $\sigma({\rm vertex})
=1$ mm and 2 mm, assuming a Gaussian distribution. With these distributions, we train boosted decision
tree classifiers, BDT and BDTD, and use the BDT(D)-response functions for the signal
($t \to W s$) and background ($t \to W b $) events. This information is used to study the background
rejection versus the signal efficiency, which would enable to achieve typically 10\% signal efficiency
and a background rejection of $10^3$. Detailed studies are done at three representative values of
the LHC centre-of-mass energies, $\sqrt{s}=7$ TeV, 10 TeV and 14 TeV. As the principal results (BDT(D)
response functions and background rejection versus signal efficiencies) are very similar for all
three energies, we present detailed results only for $\sqrt{s}=14$ TeV.   

In this exploratory study, we have made some simplifying assumptions. In particular,
 we have used PYTHIA to undertake our analysis. The cross sections for the
top pair ($pp \to t\bar{t} X$) and single top production ($pp \to t/\bar{t} X $) in PYTHIA can be
adjusted to correspond to the theoretical precision currently available. However, the distributions
and topologies, in particular for the single top (anti-top) production processes, will have to be
correctly incorporated in a realistic simulation. Likewise, we have not
attempted to define the $s$- and $b$-quark jets using a modern jet algorithm.
No attempt has been made at improving the training process by adding some more variables, like
the b-jet shapes~\cite{Aaltonen:2008de},  which are known to have some discriminating power.
We recall some of the important sources of uncertainties in our analysis: (i) predicted rates of the top quark production,
(ii) histogram shapes, (iii) integrated luminosities, (iv), efficiencies of the $b$- and $s$-quark tagging,
reflecting in our study the relative efficiencies given in Tables I and II, and the uncertainty in $m_t$,
though this effect mainly the CKM matrix element determinations form the single top (anti-top) production
processes.  All these experimental and theoretical refinements will have to be
incorporated in the analysis of the LHC data to draw quantitative conclusions. In particular, background
processes, most notably $W+jets$, $Z+jets$ will have to be considered.
Nevertheless, we have shown, in the first study of its kind, 
 that a {\it direct measurement} of $\vert V_{ts}\vert$ in top quark decays is, in principle,
 feasible at the LHC.
The simulations presented here for 14 TeV correspond to an integrated luminosity of 10 fb$^{-1}$.

 Alternative methods of determining the matrix elements $\vert V_{td}\vert$,
$\vert V_{ts}\vert$ and $\vert V_{tb}\vert$ at the LHC are based on the single
top (or anti-top) production at the LHC. One attempts to determine these matrix elements
from the cross section measurement by a simultaneous fit.
This cross section is parametrised as $\sigma (pp \to tX)= A_d \vert V_{td}\vert^2 + A_s \vert V_{ts}\vert^2 
+ A_b \vert V_{tb}\vert^2$ (and likewise for $\sigma (pp \to \bar{t}X)$), one then solves the cross-section for
 the CKM matrix elements, given the dynamical quantities $A_d$, $A_s$
and $A_b$. They, in turn, depend on estimates of the various electroweak processes
 in the single-top (or anti-top) production and on the parton distribution functions (PDFs).
Typical estimates of the
reduced cross sections at the LHC ($\sqrt{s}=14 TeV)$ are: $A_d= 766 (253)$ pb, $A_s=277 (172)$ pb,
and $A_t=150 (87)$ pb~\cite{Alwall:2006bx}, where the numbers in the parenthesis refer to the production of
 anti-top at the LHC. Based on these estimates, one expects at the LHC  $A_s/A_b \sim 2$
and $A_d/A_b \sim 5$. These ratios depend on QCD and hence will not change if the weak interactions in the
SM are modified by new physics. In the SM, one expects
 $\vert V_{ts}\vert^2/\vert V_{tb}\vert^2 \sim 1.6 \times 10^{-3}$ and   
$\vert V_{td}\vert^2/\vert V_{tb}\vert^2 \sim 6 \times 10^{-5}$. In the example of realistic beyond-the SM
physics that we are using to motivate these studies, these CKM matrix element ratios could be larger by a factor 4.
We conclude that both in the SM, and in the four generation extension of it, the cross sections
$\sigma (pp \to tX)$ and $\sigma( pp \to \bar{t}X)$ are completely
dominated by the $A_b \vert V_{tb}\vert^2$ term. Hence, 
 this proposal does not have the desired sensitivity to measure the
matrix elements  $ \vert V_{td}\vert$ and $\vert V_{ts}\vert$ at the level of theoretical interest,

 It has been recently suggested
 in~\cite{AguilarSaavedra:2010wf} that
one may improve the sensitivity to $\vert V_{td}\vert $, if one  
 refines the experimental analysis using
 the top quark rapidity distributions, which are different for the valence $d$-quark initiated
processes as opposed to the sea $b$-quark initiated processes~\cite{AguilarSaavedra:2010wf}.
While of some value in exploring $\vert V_{td}\vert $,  still 
the sensitivity of this method is   
far from the expected value of $\vert V_{td}\vert$ by an order of magnitude.  Moreover,     
 as the $s$-quark and the $b$-quark are both sea-quarks in the proton, 
the top quark rapidity distributions do not provide an improved determination of
 $\vert V_{ts}\vert$ from the single top production process. Hence, our method based on the top quark decay
characteristics to determine $\vert V_{ts}\vert$ complements the existing proposal.

Finally, we remark that the ratio of the CKM matrix elements
 $(\vert V_{td}\vert^2 + \vert V_{ts}\vert^2)/\vert V_{tb}\vert^2$, that can be obtained by measuring the
ratio $R_t$, defined in the introduction, through the number of  events with zero-, one-, and
two $b$-tags in the process $pp \to t\bar{t} X$, can be combined with the determination of the ratio
$\vert V_{ts}\vert^2/\vert V_{tb}\vert^2$ discussed here, to constrain (or measure) the quantity
$\vert V_{td}\vert^2/\vert V_{tb}\vert^2$.


Acknowledgements: We thank Karl Jakobs and Torbjorn Sjostrand for helpful discussions. Helpful
communication with Alexander Lenz is also thankfully acknowledged.

\end{document}